\documentclass[10pt]{IEEEtran}
\usepackage{amssymb}
\usepackage{amsmath}
\usepackage[dvips]{graphicx}
\usepackage{cite}

\begin{document}

\title{Direction-of-Arrival Estimation for Temporally Correlated Narrowband Signals}
\author{Farzan ~Haddadi,
        Mohammad M. Nayebi,
        and Mohammad R. Aref
\thanks{This work was supported in part by the Advanced Communication Research
Institute (ACRI), Sharif University of Technology, and in part by
the Iranian Telecommunication Research Center (ITRC) under
contract number T/500/20613. Authors are with the Department of
Electrical Engineering, Sharif University of Technology, Tehran,
Iran (e-mails: farzanhaddadi@ee.sharif.edu, Nayebi@sharif.edu, and
Aref@sharif.edu). }
\thanks{Copyright (c) 2008 IEEE. Personal use of this material is
permitted. However, permission to use this material for any other
purposes must be obtained from the IEEE by sending a request to
pubs-permissions@ieee.org.}}

\maketitle

\renewcommand{\QED}{\QEDopen}
\newtheorem{theorem}{Theorem}
\newtheorem{result}{Result}

\begin{abstract}
signal direction-of-arrival estimation using an array of sensors
has been the subject of intensive research and development during
the last two decades. Efforts have been directed to both, better
solutions for the general data model and to develop more realistic
models. So far, many authors have assumed the data to be iid
samples of a multivariate statistical model. Although this
assumption reduces the complexity of the model, it may not be true
in certain situations where signals show temporal correlation.
Some results are available on the temporally correlated signal
model in the literature. The temporally correlated stochastic
Cramer-Rao bound (CRB) has been calculated and an instrumental
variable-based method called IV-SSF is introduced. Also, it has
been shown that temporally correlated CRB is lower bounded by the
deterministic CRB. In this paper, we show that temporally
correlated CRB is also upper bounded by the stochastic iid CRB. We
investigate the effect of temporal correlation of the signals on
the best achievable performance. We also show that the IV-SSF
method is not efficient and based on an analysis of the CRB,
propose a variation in the method which boosts its performance.
Simulation results show the improved performance of the proposed
method in terms of lower bias and error variance.
\end{abstract}

\begin{keywords}
Cramer-Rao bound, temporal correlation, DOA estimation, array
signal processing.
\end{keywords}


\IEEEpeerreviewmaketitle

\section{Introduction}

\PARstart{D}{irection}-of-Arrival estimation using an array of
sensors is widely investigated in the literature in the last
decades. Many methods are developed for diverse conditions (see
e.g. \cite{krim,vanveen,schmidt, viberg_ssf,bresler}) and their
performances are presented via simulation results or analytical
calculations (for papers on theoretical analysis of the
performances of the methods see e.g. \cite{stoica_MUSIC_CRB,xu,
tichavsky,abeida}). Data models play an important role in DOA
estimation. They facilitate theoretical derivations of various
solutions by their inherent simplistic mathematical and
statistical nature. At the same time, overseeing many real-world
effects may result in modelling errors and therefore suboptimal
methods for the problem at hand. Then, there is a trade-off
between simplicity of the models and their usefulness in practice.
In the context of array signal processing, there has been
considerable research to resolve many aspects of the standard
model during the last decades. As a parallel line of research,
there also have been efforts to introduce better models which are
more application-specific or sometimes more general than the
standard model.

In the standard DOA estimation model, snapshots are assumed to be
iid or uncorrelated in time. In practice, iid assumption is
satisfied with narrowband bandpass filtering of the received
signal which gives zero time correlation in certain time delays as
ideal sampling points. However, assuming iid snapshots, places a
limitation on the applicability of the results in the real world
and also forces some practical difficulties. Assuming perfect
bandpass filtering of the signal and exact sampling on the zeros
of the correlation function, sampling rate should be reduced so
that the subsequent samples be uncorrelated. This is a technical
difficulty for environments with slowly fluctuating signals when
we need large number of samples. Some maximum likelihood methods
require the number of observations be at least equal to the number
of sensors in order to have a full rank sample covariance matrix
$\hat{\boldsymbol R}$ \cite{jaffer}, at the same time many
available DOA estimation methods are efficient only asymptotically
in number of snapshots \cite{stoica_MUSIC_CRB}. Then, assuming iid
snapshots in theory, results in longer observation times which is
not always possible due to moving targets, and may expose us to
unpredicted errors due to modelling error. Therefore, it seems
useful to accommodate time correlation of at least signals in the
model.

Recently, there have been some attempts to tackle with this
problem. In \cite{delmas}, authors consider the performance of the
spatial covariance-based methods for DOA estimation when signal
and noise are not guaranteed to be iid. They conclude that only if
noise is still uncorrelated in time, most methods are insensitive
to the time correlation of the signals. In \cite{viberg}, an
instrumental variable approach (IV-SSF) to the
direction-of-arrival estimation problem in the presence of the
time correlation of the signals is proposed. The authors improve
their previously presented method in \cite{stoicaIV} to obtain a
more reliable approach. The feature of the IV-SSF is that it does
not require any knowledge of the noise covariance matrix but its
uncorrelatedness in time. In \cite{viberg}, authors also present a
statistical performance evaluation for the IV-SSF and calculate
some performance bounds for it. In particular, they derive the CRB
for the general case of temporally correlated signals and show
that it is lower bounded by the well-known deterministic CRB
\cite{stoicaCU}. In another paper in this field \cite{stoicaCON},
authors show the asymptotic equivalence of spatial and temporal
IV-SSF methods in a unified framework.

In this paper, we study the properties of the temporally
correlated CRB. Although the effect of temporal correlation of the
signals on the conventional methods of DOA estimation is
investigated in \cite{delmas}, an important remaining question is
the role of temporal correlation of the signals in the best
achievable performance. Is it helpful or harmful and how much in
various conditions? We investigate this issue and show that it is
helpful particularly in low SNRs. We also show that the temporally
correlated $\textrm{CRB}^\textrm{cor}$ is decreasing with the
number of samples, as we expect. Then we turn to the IV-SSF to
show that it is not an efficient method of DOA estimation in the
sense that it cannot achieve the $\textrm{CRB}^\textrm{cor}$.
Using an asymptotical analysis of the $\textrm{CRB}^\textrm{cor}$,
we propose an improved version of the IV-SSF which can outperform
the existing version. At the end, simulation results confirm the
superiority of the new version in terms of lower finite sample
bias and estimation error variance.

The remaining of the paper is organized as follows: Section
\ref{sec::model} presents the data model for the temporally
correlated signals array processing. Section \ref{sec::CRB} is
dedicated to the analysis of the CRB properties and a number of
new results and comparisons. In section \ref{sec::IV-SSF} the
optimality of the IV-SSF in comparison with the
$\textrm{CRB}^\textrm{cor}$ is investigated and an improvement is
proposed. In section \ref{sec::simulation}, simulation results are
presented to show the better performance of the proposed method.
Finally, section \ref{sec::Conclusion} concludes the paper.

\emph{Notation:} \\
\begin{tabular}{ll}
$\otimes$ & Kronecker product; \\
$\odot$ & Hadamard-Schur product; \\
$\| \cdot \|$ & matrix Frobenious norm; \\
$(\cdot)^*$ & conjugate; \\
$(\cdot)^\textrm{T}$ & transpose; \\
$(\cdot)^\textrm{H}$ & conjugate transpose; \\
$\textrm{Tr}(\cdot)$ & trace; \\
$\textrm{vec} (\cdot)$ & vectorizing operator; \\
$\Re$ & real part; \\
$\delta_{ij}$ & Kronecker delta; \\
$\boldsymbol a(\theta_i)$ & array steering vector; \\
$\boldsymbol A$ & $=[ \, \boldsymbol a(\theta_1)
, \ldots , \boldsymbol a(\theta_m)]$;  steering matrix; \\
$\boldsymbol D$ & $= [\, \boldsymbol d_1 , \ldots , \boldsymbol
d_m]$; $\boldsymbol d_i = d \boldsymbol a(\theta_i) / d \theta
_i$; \\
${\boldsymbol \Pi}_{\boldsymbol A} ^ \bot$ & $\boldsymbol I -
\boldsymbol A (\boldsymbol A^\textrm{H} \boldsymbol A )^{-1}
\boldsymbol A^\textrm{H}$; orthogonal  \\ & projection on to the
null space of $\boldsymbol A$; \\
$\boldsymbol A \geq \boldsymbol B$ & $\boldsymbol A - \boldsymbol
B$ is positive semi-definite; \\
$\boldsymbol {\mathcal{P}}^{ij}$ & $ij$'th block of matrix
$\boldsymbol {\mathcal{P}}$; \\
$\boldsymbol {\mathcal{P}}_{ij}$ & $ij$'th element of matrix
$\boldsymbol {\mathcal{P}}$; \\
$\textrm{BTr}_m  (\boldsymbol {\mathcal{P}})$ & $= \sum_i
\boldsymbol {\mathcal{P}}^{ii}_{m \times m}$; block trace;\\
$\textrm{Block}_{ij} [ \boldsymbol {\mathcal{P}}^{ij} ]$ & a block
matrix
with blocks $\boldsymbol {\mathcal{P}}^{ij}$. \\
\end{tabular}

\section{Data Model}
\label{sec::model}

Let an array of $L$ sensors receive $n$ samples of the ambient
signal and noise. Signal is composed of plane waves from $m$
distant point sources with directions $\boldsymbol \theta =
[\theta_1 , ... , \theta_m]$. Spatial correlation between
different sources and temporal correlation of the signals are
permitted. Noise is assumed to be temporally white. Then, received
data can be modelled as
\begin{equation}
\label{eq::model_t} \boldsymbol x(t) = \boldsymbol A(\boldsymbol
\theta) \boldsymbol s(t) + \boldsymbol \nu(t)
\end{equation}
where $\boldsymbol s(t)_{m \times 1}$ is the sources signals and
$\boldsymbol \nu(t)_{L \times 1}$ is the sensors noise vector.
Data samples are gathered in matrix form as $\boldsymbol X_{L
\times n} = [\boldsymbol x(t_1) , \ldots , \boldsymbol x(t_n) ]$,
and the same is done for the signal sequences $\boldsymbol S_{m
\times n} = [\boldsymbol s(t_1) , \ldots , \boldsymbol s(t_n) ]$
and noise $\boldsymbol V_{L \times n} = [\boldsymbol \nu(t_1) ,
\ldots , \boldsymbol \nu(t_n) ]$. Using this notation, the model
in \eqref{eq::model_t} can be written as
\begin{equation}
\label{eq::model_matrix} \boldsymbol X = \boldsymbol A(\boldsymbol
\theta) \boldsymbol S + \boldsymbol V.
\end{equation}
Now, we consider the statistical properties of the model.
Regarding the statistics of the signal, three models can be
assumed: deterministic, iid, and temporally correlated signal
models.

\subsubsection{Deterministic Signal Model} Deterministic signal
model assumes a constant signal sequence in every realizations of
the process. Then the statistical model of received data will be
Gaussian with mean of the signal part and covariance matrix due
only to noise
\begin{eqnarray}
\label{eq::model_det} \boldsymbol x^{det}(t) \sim \mathcal{N}
\big( \boldsymbol A(\boldsymbol \theta) \boldsymbol s(t)\, ,
\boldsymbol
C \, \big) \\
\label{eq::C} \boldsymbol C = \textrm{E} \big[ \boldsymbol \nu(t)
\boldsymbol \nu ^ \textrm{H} (t) \big].
\end{eqnarray}
\subsubsection{iid Signal Model}
In the iid signal model, signal is a random process
with the same distribution in every snapshots and no correlation
exists between snapshots. Then, data model for the iid case is
\begin{equation}
\label{eq::model_iid} \boldsymbol x^{iid}(t) \sim \mathcal{N}
(\boldsymbol 0, \boldsymbol R)
\end{equation}
where
\begin{equation}
\label{eq::R_iid} \boldsymbol R = \boldsymbol {APA}^ \textrm{H} +
\boldsymbol C
\end{equation}
and
\begin{eqnarray}
\label{eq::P_iid} \boldsymbol P = \textrm{E} \big[ \boldsymbol
s(t) \boldsymbol s^ \textrm{H} (t) \big].
\end{eqnarray}
\subsubsection{Correlated Signal Model}
In the correlated signal model, noise samples are
still assumed to be spatially correlated, distributed as a
zero-mean Gaussian random vector $\boldsymbol \nu(t) \sim
\mathcal{N}(\boldsymbol 0, \boldsymbol C)$, and temporally
uncorrelated with $\textrm{E} [ \boldsymbol \nu(t_i) \boldsymbol
\nu^\textrm{H}(t_j) ] = \boldsymbol 0 \, , \, i \neq j$. Then, a
space-time distribution for noise can be defined as
\begin{equation}
\textrm{vec} (\boldsymbol V) \sim \mathcal{N} (\boldsymbol 0 ,
\boldsymbol {\mathcal{C}})
\end{equation}
\begin{equation}
\boldsymbol {\mathcal{C}} = \boldsymbol I_n \otimes \boldsymbol C.
\end{equation}
Similarly, a space-time covariance can be defined for signals. Let
$\boldsymbol {\mathcal{P}}^{ij} \triangleq \textrm{E} [
\boldsymbol s(t_i) \boldsymbol s^\textrm{H}(t_j) ]$, then
\begin{equation}
\textrm{vec}(\boldsymbol S ^{cor}) \sim \mathcal{N} (\boldsymbol 0
, \boldsymbol {\mathcal{P}})
\end{equation}
where $\boldsymbol {\mathcal{P}}_{nm \times nm} =
\textrm{Block}_{ij} [\boldsymbol {\mathcal{P}}^{ij}_{m \times m}
]$. Now, the space-time distribution of the data under correlated
signal model can be easily shown to be
\begin{equation}
\label{eq::model_st} \textrm{vec} (\boldsymbol X^{cor}) \sim
\mathcal{N} (\boldsymbol 0 , \boldsymbol {\mathcal{R}})
\end{equation}
where
\begin{equation}
\label{eq::R_st} \boldsymbol {\mathcal{R}} = \boldsymbol
{\mathcal{APA}}^\textrm{H} + \boldsymbol {\mathcal{C}}
\end{equation}
and
\begin{equation}
\label{eq::A} \boldsymbol {\mathcal{A}} = \boldsymbol I_n \otimes
\boldsymbol A(\boldsymbol \theta).
\end{equation}
Note that \eqref{eq::R_st} has a structure similar to the iid case
\eqref{eq::R_iid}. It is also assumed that $m<L$ and that the
array manifold has the property that every set of distinct
steering vectors $\{\boldsymbol a(\theta_1) , \cdots , \boldsymbol
a(\theta_L) \}$ forms a linearly independent set. Also, $
\boldsymbol a(\theta)$ is assumed to be a smooth function as it is
in real applications which means that $ \boldsymbol d(\theta)$
exists. These assumptions pave the way for the estimation problem
at hand to be identifiable.

\section{Analysis of the CRB}
\label{sec::CRB}

In this section, we present new results and analyzes on the
temporally correlated CRB to give more insight on the estimation
problem. We are specially interested in the role of temporal
correlation of the signals in the direction estimation problem in
comparison with the uncorrelated signal case or iid signals.

CRB is a lower bound on the performance of any unbiased estimation
method in terms of error variance. Consider a random vector
distributed as $\boldsymbol y(t) \sim f(\boldsymbol \psi)$, where
$\boldsymbol \psi$ is the vector of possibly unknown parameters of
the distribution. Given samples of $\boldsymbol y(t)$, an unbiased
estimator $\hat{\boldsymbol \psi}$ satisfies $\textrm{E}
[\hat{\boldsymbol \psi}] = \boldsymbol \psi$ and its error
variance is lower bounded by the CRB as
\begin{equation}
\label{eq::CRB_def} \textrm{E} \big[ (\hat{\boldsymbol \psi} -
\boldsymbol \psi) (\hat{\boldsymbol \psi} - \boldsymbol \psi)
^\textrm{H} \big] \geq \textrm{CRB}_{\boldsymbol \psi ,
\boldsymbol \psi}.
\end{equation}
CRB is an inherent property of the statistical model of the data
and can be calculated directly from $f(\boldsymbol \psi)$. The
importance of the CRB also comes from the fact that there exist
estimators that at least asymptotically attain the CRB such as the
maximum likelihood estimator. In the following, CRB for three
discussed signal models are repeated for the sake of reference. In
the deterministic signal model of \eqref{eq::model_det} and
\eqref{eq::C} we have \cite{stoicaCU}
\begin{eqnarray}
\label{eq::CRB_det} \textrm{CRB}_{\boldsymbol \theta , \boldsymbol
\theta} ^ \textrm{det} = \frac{1}{2n} \Big[ \Re \big(\boldsymbol
D^\textrm{H} \boldsymbol C ^{-\frac{1}{2}} {\boldsymbol
\Pi}_0^\bot \boldsymbol C ^{-\frac{1}{2}} \boldsymbol D \big)
\odot \boldsymbol P^\textrm{T} \Big] ^ {-1}
\end{eqnarray}
where
\begin{equation}
{\boldsymbol \Pi}_0^\bot = {\boldsymbol \Pi}_{\boldsymbol C
^{-\frac{1}{2}} \boldsymbol A}^\bot.
\end{equation}
In the iid model of \eqref{eq::model_iid} and \eqref{eq::R_iid}
the $\textrm{CRB} ^ {\textrm{iid}}$ is \cite{stoicaCU, stoicaCRB}
\begin{eqnarray}
\label{eq::CRB_iid} \textrm{CRB}_{\boldsymbol \theta , \boldsymbol
\theta} ^ \textrm{iid} = \frac{1}{2n} \Big[ \Re \big(\boldsymbol
D^\textrm{H} \boldsymbol C ^{-\frac{1}{2}} {\boldsymbol
\Pi}_0^\bot \boldsymbol C ^{-\frac{1}{2}} \boldsymbol D \big)
\quad \quad \nonumber \\
\odot \big( \boldsymbol {PA}^\textrm{H} \boldsymbol R ^ {-1}
\boldsymbol {A \, P} \big)^{\textrm{T}} \Big] ^ {-1}
\end{eqnarray}
and in the temporally correlated signal model of
\eqref{eq::model_st}, \eqref{eq::R_st}, and \eqref{eq::A} the
$\textrm{CRB} ^ {\textrm{cor}}$ is \cite[eq. 112]{viberg}
\begin{eqnarray}
\label{eq::CRB_cor} \textrm{CRB}_{\boldsymbol \theta , \boldsymbol
\theta} ^ {\textrm{cor}} = \frac{1}{2} \Big[ \Re \big(\boldsymbol
D^\textrm{H} \boldsymbol C ^{-\frac{1}{2}} {\boldsymbol
\Pi}_0^\bot \boldsymbol
C ^{-\frac{1}{2}} \boldsymbol D \big) \qquad \quad \nonumber \\
\odot \, \textrm{BTr}_m ( \boldsymbol {\mathcal{PG}}^\textrm{H}
\boldsymbol {\, \mathcal{R'}} ^ {-1} \boldsymbol {\! \mathcal{G \,
P}} ) ^ \textrm{T} \Big] ^ {-1}.
\end{eqnarray}
where
\begin{equation}
\boldsymbol {\mathcal G} = \boldsymbol {\mathcal{C}} ^
{-\frac{1}{2}} \boldsymbol {\mathcal{A}}
\end{equation}
\begin{equation}
\boldsymbol {\mathcal{R'}} = \boldsymbol {\mathcal {GPG}} ^
\textrm{H} + \boldsymbol {\mathcal{I}} = \boldsymbol {\mathcal{C}}
^ {-\frac{1}{2}} \, \boldsymbol {\mathcal R} \,  \boldsymbol
{\mathcal{C}} ^ {-\frac{1}{2}}.
\end{equation}
where $\boldsymbol {\mathcal{I}} = \boldsymbol I_n \otimes
\boldsymbol I_L$. It is noteworthy that the expression of the
$\textrm{CRB} ^ {\textrm{cor}}$ in \cite{viberg} is slightly
different from \eqref{eq::CRB_cor} in that it contains a factor of
$\frac{1}{n}$ in the right-hand-side. This is a direct consequence
of the difference in the definition of the CRB between
\eqref{eq::CRB_def} and what is in \cite{viberg}. As a
confirmation for the form of $\textrm{CRB} ^ {\textrm{cor}}$ in
\eqref{eq::CRB_cor}, consider the case of iid signals where the
space-time matrices turn out to be block-diagonal. then a factor
of $\frac{1}{n}$ appears in the right-hand-side to reduce the
$\textrm{CRB} ^ {\textrm{cor}}$ in \eqref{eq::CRB_cor} to the
$\textrm{CRB} ^ {\textrm{iid}}$ in \eqref{eq::CRB_iid}.

To proceed further, we first give a simplified form of the
$\textrm{CRB} ^ {\textrm{cor}}$ in \eqref{eq::CRB_cor}. Note that
\begin{equation}
\boldsymbol {\mathcal{G}}^\textrm{H} \boldsymbol {\, \mathcal{R'}}
^ {-1} \boldsymbol {\! \mathcal{G}} = \boldsymbol
{\mathcal{A}}^\textrm{H} \boldsymbol {\, \mathcal{R}} ^ {-1}
\boldsymbol {\! \mathcal{A}}
\end{equation}
then the $\textrm{CRB} ^ {\textrm{cor}}$ can be written as
\begin{eqnarray}
\label{eq::CRB_cor_simp} \textrm{CRB}_{\boldsymbol \theta ,
\boldsymbol \theta} ^ {\textrm{cor}} = \frac{1}{2} \Big[ \Re
\big(\boldsymbol D^\textrm{H} \boldsymbol C ^{-\frac{1}{2}}
{\boldsymbol \Pi}_0^\bot \boldsymbol
C ^{-\frac{1}{2}} \boldsymbol D \big) \qquad \quad \nonumber \\
\odot \, \textrm{BTr}_m ( \boldsymbol {\mathcal{PA}}^\textrm{H}
\boldsymbol {\, \mathcal{R}} ^ {-1} \boldsymbol {\! \mathcal{A \,
P}} ) ^ \textrm{T} \Big] ^ {-1}.
\end{eqnarray}
Now, we are ready to state our first result on the comparison of
the CRB in the signal models introduced.

\begin{theorem}
\label{thrm::bound} For the deterministic signal model in
\eqref{eq::model_det} and \eqref{eq::C}, iid signal model in
\eqref{eq::model_iid} and \eqref{eq::R_iid}, and the temporally
correlated signal model in \eqref{eq::model_st} and
\eqref{eq::R_st}, $\textrm{CRB} ^ {\textrm{cor}}$ is upper bounded
by $\textrm{CRB} ^ {\textrm{iid}}$ and lower bounded by
$\textrm{CRB} ^ {\textrm{det}}$
\begin{equation}
\label{eq::thrm_CRB} \textrm{CRB} ^ {\textrm{det}} \leq
\textrm{CRB} ^ {\textrm{cor}} \leq \textrm{CRB} ^ {\textrm{iid}}.
\end{equation}
\end{theorem}

\begin{proof}
The left side inequality is proved in \cite{viberg}, therefore we
give a proof for the right side inequality. We make use of a form
of the Woodburry identity \cite{bienvenu, householder} which
states that
\begin{eqnarray}
\label{eq::woodburry} (\boldsymbol A + \boldsymbol {BCD})^{-1} =
\qquad \qquad \qquad \qquad \qquad \qquad \nonumber \\ \boldsymbol
A^{-1} - \boldsymbol A^{-1} \boldsymbol B (\boldsymbol I +
\boldsymbol {CDA} ^{-1} \boldsymbol B)^{-1} \boldsymbol {CDA}
^{-1}
\end{eqnarray}
which is essentially another form of the matrix inversion lemma.
Now we expand the space-time matrix in the block trace of
\eqref{eq::CRB_cor_simp} substituting from \eqref{eq::R_st} and
making use of \eqref{eq::woodburry} to have
\begin{eqnarray}
\label{eq::pr_1} \boldsymbol {\mathcal{PA}}^\textrm{H} \boldsymbol
{\, \mathcal{R}} ^ {-1} \boldsymbol {\! \mathcal{A \, P}} =
\boldsymbol {\mathcal {PA}} ^\textrm{H} \big[  \boldsymbol
{\mathcal C}^{-1} - \nonumber
\qquad \quad \qquad \qquad \\
\boldsymbol {\mathcal C}^{-1} \boldsymbol {\mathcal A} \, (
\boldsymbol {\mathcal I} + \boldsymbol {\mathcal {PA}}^\textrm{H}
\boldsymbol {\mathcal C}^{-1} \boldsymbol {\mathcal A} )^{-1} \,
\boldsymbol {\mathcal {PA}}^\textrm{H} \boldsymbol {\mathcal
C}^{-1} \big] \boldsymbol {\mathcal {AP}} = \nonumber \\
\boldsymbol {\mathcal {PA}}^\textrm{H} \boldsymbol {\mathcal
C}^{-1} \boldsymbol {\mathcal {AP}} - \boldsymbol {\mathcal
{PA}}^\textrm{H} \boldsymbol {\mathcal C}^{-1} \boldsymbol
{\mathcal A} (\boldsymbol {\mathcal I} +   \qquad
\nonumber \\
\boldsymbol {\mathcal {PA}}^\textrm{H} \boldsymbol {\mathcal
C}^{-1} \boldsymbol {\mathcal A} )^{-1} \boldsymbol {\mathcal
{PA}}^\textrm{H} \boldsymbol {\mathcal C}^{-1} \boldsymbol
{\mathcal {AP}}.
\end{eqnarray}
We factor the common terms of \eqref{eq::pr_1} from right side to
get
\begin{eqnarray}
\label{eq::pr_2} \big[ \boldsymbol {\mathcal I} - \boldsymbol
{\mathcal {PA}}^\textrm{H} \boldsymbol {\mathcal C}^{-1}
\boldsymbol {\mathcal A} \, (
\boldsymbol {\mathcal I} + \qquad \qquad \qquad \nonumber \\
\boldsymbol {\mathcal {PA}}^\textrm{H} \boldsymbol {\mathcal
C}^{-1} \boldsymbol {\mathcal A} )^{-1} \big] \boldsymbol
{\mathcal {PA}}^\textrm{H} \boldsymbol {\mathcal C}^{-1}
\boldsymbol {\mathcal {AP}}.
\end{eqnarray}
Add and subtract an $\boldsymbol {\mathcal I}$ to the $\boldsymbol
{\mathcal {PA}}^\textrm{H} \boldsymbol {\mathcal C}^{-1}
\boldsymbol {\mathcal A}$ in the bracket in \eqref{eq::pr_2} to
get
\begin{equation}
\label{eq::pr_3}  ( \boldsymbol {\mathcal I} +  \boldsymbol
{\mathcal {PA}}^\textrm{H} \boldsymbol {\mathcal C}^{-1}
\boldsymbol {\mathcal A} )^{-1} \boldsymbol {\mathcal
{PA}}^\textrm{H} \boldsymbol {\mathcal C}^{-1} \boldsymbol
{\mathcal {AP}}.
\end{equation}
Do the same for the term outside the bracket in \eqref{eq::pr_3}
in the following form
\begin{eqnarray}
\label{eq::pr_4} ( \boldsymbol {\mathcal I} +  \boldsymbol
{\mathcal {PA}}^\textrm{H} \boldsymbol {\mathcal C}^{-1}
\boldsymbol {\mathcal A} )^{-1} ( \boldsymbol {\mathcal I} +
\boldsymbol {\mathcal {PA}}^\textrm{H} \boldsymbol {\mathcal
C}^{-1} \boldsymbol {\mathcal A} - \boldsymbol {\mathcal I})
\boldsymbol {\mathcal P}.
\end{eqnarray}
As a useful result, we arrive to the following equality
simplifying \eqref{eq::pr_4}
\begin{eqnarray}
\label{eq::pr_5} \boldsymbol {\mathcal{PA}}^\textrm{H} \boldsymbol
{\, \mathcal{R}} ^ {-1} \boldsymbol {\! \mathcal{A \, P}} =
\boldsymbol {\mathcal P} - ( \boldsymbol {\mathcal P}^{-1} +
\boldsymbol {\mathcal A}^\textrm{H} \boldsymbol {\mathcal C}^{-1}
\boldsymbol {\mathcal A} )^{-1}.
\end{eqnarray}
Note that the matrix in parenthesis is positive semi-definite,
then we will have
\begin{eqnarray}
\label{eq::pr_6} \boldsymbol {\mathcal{PA}}^\textrm{H} \boldsymbol
{\, \mathcal{R}} ^ {-1} \boldsymbol {\! \mathcal{A \, P}} \leq
\boldsymbol {\mathcal P}
\end{eqnarray}
which is another proof for the left side inequality in
\eqref{eq::thrm_CRB} (see \cite{viberg}). Now, we proceed to prove
the right side inequality in \eqref{eq::thrm_CRB}. define
\begin{equation}
\boldsymbol {\mathcal P}_d \triangleq \textrm{Block}_{ij} [
\boldsymbol {\mathcal P}^{ij} \delta_{ij} ]
\end{equation}
which is the block-diagonalized version of the signal space-time
covariance matrix. Note that block-diagonal space-time covariance
matrix for signals $\boldsymbol {\mathcal P}_d$, represents the
temporally uncorrelated signal model (more general than iid signal
model), while $\boldsymbol {\mathcal P}$ represents the temporally
correlated signal model. Now, assume a matrix $\boldsymbol
{\mathcal F}$ and its block-diagonalized version $\boldsymbol
{\mathcal F}_d$. It is well known that
\begin{equation}
\label{eq::diag_ineq} (\boldsymbol {\mathcal F}^{-1})^{ii} \geq
(\boldsymbol {\mathcal F}^{ii})^{-1} = (\boldsymbol {\mathcal
F}_d^{ii})^{-1} = (\boldsymbol {\mathcal F}_d^{-1})^{ii}.
\end{equation}
Using the inequality in \eqref{eq::diag_ineq}, besides the fact
that $\boldsymbol {\mathcal A}^\textrm{H} \boldsymbol {\mathcal
C}^{-1} \boldsymbol {\mathcal A}$ is block-diagonal, we will have
\begin{eqnarray}
\label{eq::pr_7}  \boldsymbol {\mathcal A}^\textrm{H} \boldsymbol
{\mathcal C}^{-1} \boldsymbol {\mathcal A} - \big(\boldsymbol
{\mathcal P} + (\boldsymbol {\mathcal A}^\textrm{H} \boldsymbol
{\mathcal C}^{-1} \boldsymbol {\mathcal
A})^{-1} \big)^{-1}  \leq  \qquad \nonumber \\
\boldsymbol {\mathcal A}^\textrm{H} \boldsymbol {\mathcal C}^{-1}
\boldsymbol {\mathcal A} - \big(\boldsymbol {\mathcal P}_d +
(\boldsymbol {\mathcal A}^\textrm{H} \boldsymbol {\mathcal C}^{-1}
\boldsymbol {\mathcal A})^{-1} \big)^{-1}
\end{eqnarray}
Now, we multiply $\boldsymbol {\mathcal B} \triangleq (\boldsymbol
{\mathcal A}^\textrm{H} \boldsymbol {\mathcal C}^{-1} \boldsymbol
{\mathcal A})^{-1}$ from right and left of both sides of the
inequality to get
\begin{eqnarray}
\label{eq::pr_8}  \boldsymbol {\mathcal B} - \boldsymbol {\mathcal
B} \big(\boldsymbol {\mathcal P} + \boldsymbol {\mathcal B}
\big)^{-1} \boldsymbol {\mathcal B} \leq  \boldsymbol {\mathcal B}
- \boldsymbol {\mathcal B} \big(\boldsymbol {\mathcal P}_d +
\boldsymbol {\mathcal B} \big)^{-1} \boldsymbol {\mathcal B}.
\end{eqnarray}
The above expressions can be simplified using matrix inversion
lemma to give
\begin{eqnarray}
\label{eq::pr_9}  \big( \boldsymbol {\mathcal P}^{-1} +
\boldsymbol {\mathcal A}^\textrm{H} \boldsymbol {\mathcal C}^{-1}
\boldsymbol {\mathcal A} \big)^{-1} \leq  \big(  \boldsymbol
{\mathcal P}_d^{-1} + \boldsymbol {\mathcal A}^\textrm{H}
\boldsymbol {\mathcal C}^{-1} \boldsymbol {\mathcal A} \big)^{-1}.
\end{eqnarray}
Applying the block trace operator and adding a common term results
in
\begin{eqnarray}
\label{eq::pr_10}  \textrm{BTr}_m \big[ \boldsymbol {\mathcal P} -
\big(\boldsymbol {\mathcal P}^{-1} + \boldsymbol {\mathcal
A}^\textrm{H} \boldsymbol {\mathcal C}^{-1} \boldsymbol {\mathcal
A} \big)^{-1} \big] \geq \qquad \qquad \nonumber \\
\textrm{BTr}_m \big[ \boldsymbol {\mathcal P}_d - \big(\boldsymbol
{\mathcal P}_d^{-1} + \boldsymbol {\mathcal A}^\textrm{H}
\boldsymbol {\mathcal C}^{-1} \boldsymbol {\mathcal A} \big)^{-1}
\big].
\end{eqnarray}
According to \eqref{eq::pr_5}, the inequality in \eqref{eq::pr_10}
implies that
\begin{eqnarray}
\label{eq::pr_11}  \textrm{BTr}_m \big[ \boldsymbol
{\mathcal{PA}}^\textrm{H} \boldsymbol{\, \mathcal{R}} ^ {-1}
\boldsymbol {\! \mathcal{A \, P}} \big] \geq \qquad \qquad
\nonumber \\
\textrm{BTr}_m \big[ \boldsymbol {\mathcal P}_d \boldsymbol
{\mathcal A}^\textrm{H} \boldsymbol {\, \mathcal{R}}_d ^ {-1}
\boldsymbol {\! \mathcal{A \, P}}_d \big].
\end{eqnarray}
which in fact completes the proof of \eqref{eq::thrm_CRB} showing
that
\begin{equation}
\label{eq::CRB_Pd_ineq} \textrm{CRB} ^ {\textrm{cor}} (\boldsymbol
{\mathcal P}) \leq \textrm{CRB} ^ {\textrm{cor}} (\boldsymbol
{\mathcal P}_d).
\end{equation}
\end{proof}

The upper bound and lower bound on the $\textrm{CRB} ^
{\textrm{cor}}$ presented in \eqref{eq::thrm_CRB} is very
insightful to the estimation problem at hand. It implies that the
existence of a temporal correlation in the signals improve the
best attainable performance of estimation. The comparison made in
theorem \ref{thrm::bound} is conditioned on the specific spatial
covariance matrix of the sources in each sample i.e. with the same
diagonal blocks of the space-time signal covariance matrix. It
shows that adding a nondiagonal covariance block improves the
performance since it simplifies the extraction of the signal part
from received data. Most spatial covariance-based DOA estimation
methods rely on the different spatial characteristics of the
signal and noise (signals are point sources of radiation while
noise is uniformly distributed in the space or at least is spread
via large areas). This is not the case in the temporally
correlated signal model where there is a particular difference
between signal and noise in that one is temporally correlated and
the other is temporally uncorrelated. This increased distance of
the signal model and noise model improves the CRB and the
performance of the methods using it. As a result, we can see that
the $\textrm{CRB} ^ {\textrm{cor}}$ is lower than the
$\textrm{CRB} ^ {\textrm{iid}}$ which implies better performance
when noise and signal have different temporal characteristics.
Another explanation for the inequalities presented in
\eqref{eq::thrm_CRB} comes from the degree of predictability of
the signal. It is obvious that in the deterministic signal model
we have a statistically constant signal which is fully
predictable. Therefore, the performance is best in the
deterministic signal model. In the correlated signal model,
signals are stochastic in nature and vary in each realization,
which makes the signals less predictable. Though, the existence of
the temporal correlation of the signals offers a limited
possibility for coarse signal prediction and extraction from noise
which places the $\textrm{CRB} ^ {\textrm{cor}}$ lower than the
completely uncorrelated case of $\textrm{CRB} ^ {\textrm{iid}}$.

Although we have confined the $\textrm{CRB} ^ {\textrm{cor}}$
between $\textrm{CRB} ^ {\textrm{det}}$ and $\textrm{CRB} ^
{\textrm{iid}}$ in theorem \ref{thrm::bound}, we are interested to
more exactly specify the behavior of the $\textrm{CRB} ^
{\textrm{cor}}$ in different situations. This helps us to get more
insight to the role of temporal correlation of the signals in DOA
estimation. Therefore we consider approximations of the
$\textrm{CRB} ^ {\textrm{cor}}$ in different situations in temrs
of signal to noise ratio (SNR) in the following theorem.

\begin{theorem}
\label{thrm::snr} In the high SNR condition, temporal correlation
of the signals makes no improvement on the uncorrelated signal
model while in the low SNR condition, the contribution of zero-lag
and nonzero-lag covariances are the same, i.e. nonzero-lag
covariances improve the CRB.
\begin{eqnarray}
\label{eq::thrm_snr_high} \textrm{SNR} \gg 1   : \quad
\textrm{CRB} ^ {\textrm{cor}} \simeq \textrm{CRB} ^ {\textrm{iid}} \\
\label{eq::thrm_snr_low} \textrm{SNR} \ll 1   : \quad \textrm{CRB}
^ {\textrm{cor}} < \textrm{CRB} ^ {\textrm{iid}}
\end{eqnarray}
\end{theorem}

\begin{proof}
Consider the high SNR condition. We make use of the following
easily checked approximation for any appropriately sized matrices
$\boldsymbol B$ and $\boldsymbol \Delta$ if $ \| \boldsymbol
\Delta \| \leq \| \boldsymbol B \|$
\begin{eqnarray}
\label{eq::inv_approx} (\boldsymbol B + \boldsymbol \Delta)^{-1}
\simeq \boldsymbol B^{-1} -  \qquad \qquad \qquad \qquad \qquad
\nonumber \\
\boldsymbol B^{-1} \boldsymbol \Delta \boldsymbol B^{-1} +
\boldsymbol B^{-1} \boldsymbol \Delta \boldsymbol B^{-1}
\boldsymbol \Delta \boldsymbol B^{-1} - \cdots
\end{eqnarray}
Now we can expand the block trace in $\textrm{CRB} ^
{\textrm{cor}}$ as follows
\begin{eqnarray}
\label{eq::pr_12}  \textrm{BTr}_m \big[ \boldsymbol
{\mathcal{PA}}^\textrm{H} \boldsymbol{\, \mathcal{R}} ^ {-1}
\boldsymbol {\! \mathcal{A \, P}} \big] = \qquad \qquad \qquad
\qquad \nonumber \\
\textrm{BTr}_m \big[ \boldsymbol {\mathcal P} \boldsymbol
{\mathcal A}^\textrm{H} (\boldsymbol {\mathcal{C}} + \boldsymbol
{\mathcal{APA}}^\textrm{H} )^{-1} \boldsymbol {\! \mathcal{A \,
P}} \big].
\end{eqnarray}
We use matrix inversion lemma for the inverse in the block trace
of \eqref{eq::pr_12} to get
\begin{eqnarray}
\label{eq::pr_13} \textrm{BTr}_m \big[ \boldsymbol {\mathcal P}
\boldsymbol {\mathcal A}^\textrm{H} \big( \boldsymbol
{\mathcal{C}}^{-1} - \boldsymbol {\mathcal{C}}^{-1} \boldsymbol
{\mathcal{A}} ( \boldsymbol {\mathcal P}^{-1} +
\qquad \nonumber \\
\boldsymbol {\mathcal A}^\textrm{H} \boldsymbol {\mathcal C}^{-1}
\boldsymbol {\mathcal A} )^{-1} \boldsymbol
{\mathcal{A}}^\textrm{H} \boldsymbol {\mathcal{C}}^{-1} \big)
\boldsymbol { \mathcal{A \, P}} \big].
\end{eqnarray}
Now we use the approximation in \eqref{eq::inv_approx} up to the
third term. Note that the high SNR assumption guarantees that $\|
\boldsymbol {\mathcal P}^{-1} \| \leq \| \boldsymbol {\mathcal
A}^\textrm{H} \boldsymbol {\mathcal C}^{-1} \boldsymbol {\mathcal
A} \|$. After some calculations we get
\begin{eqnarray}
\label{eq::pr_14}  \textrm{BTr}_m \big[ \boldsymbol
{\mathcal{PA}}^\textrm{H} \boldsymbol{\, \mathcal{R}} ^ {-1}
\boldsymbol {\! \mathcal{A \, P}} \big] \simeq \qquad \qquad
\qquad \qquad \nonumber \\
\textrm{BTr}_m \big[ \boldsymbol {\mathcal P} - (\boldsymbol
{\mathcal A}^\textrm{H} \boldsymbol {\mathcal C}^{-1} \boldsymbol
{\mathcal A} )^{-1}  \big]
\end{eqnarray}
in the high SNR region and to the second order of approximation.
The approximation in \eqref{eq::pr_14} asserts that in high SNR
condition, only block trace of the signal space-time covariance
matrix contribute to the Cramer-Rao bound, hence we can conclude
that in this case, the temporal correlation of the signals
(represented by nondiagonal blocks of $\boldsymbol {\mathcal P}$),
do not improve the best achievable performance of the DOA
estimation and \eqref{eq::thrm_snr_high} follows.

In the very low SNR region, we use the approximation $\boldsymbol
{\mathcal R} \simeq \boldsymbol {\mathcal C}$. Then expanding the
block trace in the CRB gives
\begin{eqnarray}
\label{eq::pr_16}  \textrm{BTr}_m \big[ \boldsymbol
{\mathcal{PA}}^\textrm{H} \boldsymbol{\, \mathcal{R}} ^ {-1}
\boldsymbol {\! \mathcal{A \, P}} \big] \simeq \qquad
\nonumber \\
\sum_{i,r} \, (\boldsymbol {A \mathcal P}^{ri})^\textrm{H}
\boldsymbol C^{-1} (\boldsymbol {A \mathcal P}^{ri}).
\end{eqnarray}
We can see from \eqref{eq::pr_16} that every blocks of
$\boldsymbol {\mathcal P}$ contribute the same to the
$\textrm{CRB} ^ {\textrm{cor}}$ in the situation of very low SNR.
Since each term in the summation of \eqref{eq::pr_16} is positive
semi-definite, then each nondiagonal block of $\boldsymbol
{\mathcal P}$ improves the $\textrm{CRB} ^ {\textrm{cor}}$ making
the distance from $\textrm{CRB} ^ {\textrm{iid}}$ larger which
implies \eqref{eq::thrm_snr_low}. It is noteworthy that
improvement of the CRB does not mean the performance improvement
of the conventional DOA estimation methods, rather it clarifies
the existence of methods that can achieve better performances
through making use of the temporal characteristics of the signals
and noise.
\end{proof}

Now, after we considered the role of the temporal correlation of
the signals in the best achievable performance of DOA estimation,
we turn to an assumed characteristic of the CRB. The CRB usually
decreases with increased amount of data. This is obvious in the
iid signal models where the CRB for $n$ data samples is
$\frac{1}{n}$ of the CRB for one sample. Though, this is not very
clear for the temporally correlated signal model where the
dependence of the $\textrm{CRB} ^ {\textrm{cor}}$ on $n$ is
embedded in the size of the space-time matrices.

\begin{theorem}
\label{thrm::n} $\textrm{CRB} ^ {\textrm{cor}}$ is decreasing with
increasing $n$.
\begin{equation}
\label{eq::thrm_CRB_n} \textrm{CRB} ^ {\textrm{cor}} (n+1) <
\textrm{CRB} ^ {\textrm{cor}} (n).
\end{equation}
\end{theorem}

\begin{proof}
Applying matrix inversion lemma, we can show that
\begin{equation}
\label{eq::pr_18} \boldsymbol {\mathcal A}^\textrm{H} \boldsymbol
{\mathcal R}^{-1} \boldsymbol {\mathcal A} = \big( \boldsymbol
{\mathcal P} + ( \boldsymbol {\mathcal A}^\textrm{H} \boldsymbol
{\mathcal C}^{-1} \boldsymbol {\mathcal A})^{-1} \big)^{-1}.
\end{equation}
Assuming $ \boldsymbol {\mathcal{P}}_{n+1}$ as the
block-diagonally augmented version of $ \boldsymbol
{\mathcal{P}}_{n}$ and using \eqref {eq::pr_18}, we will have
\begin{equation}
\label{eq::pr_20} \textrm{CRB} ^ {\textrm{cor}} \Big( \left[ \!\!
\begin{array}{l} \boldsymbol {\mathcal P}_n
\\ \qquad \boldsymbol P_{n+1} \end{array} \!\! \right] \Big)
\leq \textrm{CRB} ^ {\textrm{cor}} (\boldsymbol {\mathcal P}_n).
\end{equation}
since
\begin{eqnarray}
\label{eq::pr_17} \textrm{BTr}_m [ \boldsymbol {\mathcal P}_{n+1}
\boldsymbol {\mathcal A}^\textrm{H}_{n+1} \boldsymbol{\,
\mathcal{R}} ^ {-1} _{n+1} \boldsymbol { \mathcal A}_{n+1}
\boldsymbol {\mathcal P}_{n+1} \big] = \qquad \qquad \nonumber \\
\textrm{BTr}_m \big[ \boldsymbol {\mathcal P}_n ( \boldsymbol
{\mathcal P}_n + (\boldsymbol {\mathcal A}_n^\textrm{H}
\boldsymbol {\mathcal C}^{-1}_n \boldsymbol {\mathcal
A}_n)^{-1})^{-1} \boldsymbol {\mathcal P}_n \big] + \qquad
\nonumber \\
\boldsymbol P_{n+1} \big( \boldsymbol P_{n+1} + (\boldsymbol
A^\textrm{H} \boldsymbol C^{-1} \boldsymbol A)^{-1} \big)^{-1}
\boldsymbol P_{n+1}
\end{eqnarray}
and the second term is positive semi-definite. Following the same
steps as in \eqref{eq::pr_7} to \eqref{eq::CRB_Pd_ineq}, we can
also show the following inequality which completes the proof.
\begin{eqnarray}
\label{eq::pr_21}\textrm{CRB} ^ {\textrm{cor}} \Big( \left[ \!\!
\begin{array}{lc} \boldsymbol {\mathcal P}_n & \boldsymbol Q ^\textrm{H}
\\ \boldsymbol Q  & \boldsymbol P_{n+1} \end{array} \!\! \right] \Big)
\leq \textrm{CRB} ^ {\textrm{cor}} \Big( \left[ \!\!
\begin{array}{l} \boldsymbol {\mathcal P}_n
\\ \qquad \boldsymbol P_{n+1} \end{array} \!\! \right] \Big)
\end{eqnarray}
Note that, \eqref{eq::pr_21} is a generalization of the R.H.S.
inequality in \eqref{eq::thrm_CRB}.
\end{proof}

In this section, we investigated the general properties of the
direction-of-arrival estimation problem in the presence of the
temporally correlated signals. We performed this via Cramer-Rao
bound analysis and characterization. In the next section we turn
to the practical methods to accomplish DOA estimation under
temporally correlated signal model.

\section{Suboptimality of the IV-SSF Method}
\label{sec::IV-SSF}

We have considered the best achievable performance in the
temporally correlated signal model. In particular, we found that
it is possible to improve the performance of the conventional
spatial covariance-based methods by devising new methods that can
exploit the temporal correlation of the signals. However, the
$\textrm{CRB} ^ {\textrm{cor}}$ has been calculated under the
assumption of known noise spatial covariance matrix. When this is
not true, we do not expect any method to reach the $\textrm{CRB} ^
{\textrm{cor}}$ in performance. In the situation of unknown noise
spatial covariance matrix, the instrumental variable subspace
fitting (IV-SSF) method for direction-of-arrival estimation has
been proposed in \cite{viberg}. The method is based on the
instrumental variable approach. It makes an instrumental variables
vector $\boldsymbol \phi(t)$ for each data sample $\boldsymbol
x(t)$ in such a way that the cross-covariance of $\boldsymbol
\phi(t)$ and $\boldsymbol x(t)$ does not contain the unknown noise
covariance matrix. Then, a sample cross-covariance can be used to
extract the signal subspace and parameters of interest. Note that
the signal temporal correlation leave a room for the
multiplication of noncontemporary data to contain information
about the directions of arrival. The instrumental variables vector
for each data sample $\boldsymbol x(t)$ is formed as
\begin{equation}
\boldsymbol \phi(t) \triangleq \left[ \!\! \begin{array}{c}
\boldsymbol x(t-1) \\ \vdots \\ \boldsymbol x(t-M) \end{array}
\!\! \right]
\end{equation}
where $M$ is a user defined integer determining the degree of
complexity and hence the performance of the method. In general,
larger $M$ should result in better estimates, although simulation
results show increased bias and decreased error variance when $M$
becomes larger. Let the cross-covariance of the signals at time
lag $k$ be defined as
\begin{equation}
\boldsymbol P_k \triangleq \textrm{E} [\boldsymbol s(t-k)
\boldsymbol s^\textrm{H}(t)]
\end{equation}
and define for convenience
\begin{equation}
\boldsymbol {\mathcal J} = \left[ \! \! \begin{array}{c} \boldsymbol P_1 \\
\vdots \\ \boldsymbol P_M \end{array} \!\!\! \right]
\end{equation}
then the cross-covariance of the received data and the
corresponding instrumental variable will be
\begin{equation}
\boldsymbol \Sigma \triangleq \textrm{E} [\boldsymbol \phi(t)
\boldsymbol x^\textrm{H}(t)] = (\boldsymbol I_M \otimes
\boldsymbol A) \, \boldsymbol {\mathcal J} \boldsymbol
A^\textrm{H}
\end{equation}
which is independent of the unknown noise spatial covariance
matrix $\boldsymbol C$. Also define the instrumental variable
covariance matrix as
\begin{equation}
\boldsymbol \Phi \triangleq \textrm{E} [\boldsymbol \phi(t)
\boldsymbol \phi^\textrm{H}(t)].
\end{equation}
The estimates of the DOAs in IV-SSF method are obtained in the
following steps: choose $M>1$ and compute the sample estimates
\begin{equation}
\hat {\boldsymbol \Sigma} = \frac{1}{n-M} \! \! \! \sum_{\; \;
t=M+1}^n \! \! \! \boldsymbol \phi(t) \boldsymbol x^\textrm{H}(t)
\end{equation}
\begin{equation}
\hat {\boldsymbol \Phi} = \frac{1}{n-M} \! \! \! \sum_{\; \;
t=M+1}^{n} \! \! \! \boldsymbol \phi(t) \boldsymbol \phi
^\textrm{H}(t).
\end{equation}
Next, extract $\hat{ \boldsymbol R}_0$ from $\hat {\boldsymbol
\Phi}$, as one of the $m \times m$ diagonal blocks. The estimates
of the parameters are the minimizer of the following criterion
function:
\begin{equation}
\hat{\boldsymbol \theta} = \arg \; \min _ {\boldsymbol \theta} \,
\textrm{Tr} \, \Big( \hat{\boldsymbol \Pi}_0^\bot \hat{\boldsymbol
R}_0^{-\frac{1}{2}} \hat{\boldsymbol V}_s \hat{\boldsymbol
\Phi}_s^2 \hat{\boldsymbol V}_s^\textrm{H} \hat{ \boldsymbol R}_0
^{-\frac {1}{2}}  \Big)
\end{equation}
where
\begin{equation}
\hat{\boldsymbol \Pi}_0 ^\bot = \boldsymbol I - \hat{\boldsymbol
R}_0 ^{-\frac{1}{2}} \boldsymbol A \big( \boldsymbol A
^{\textrm{H}} \hat{\boldsymbol R}_0 ^{-1} \boldsymbol A \big)^{-1}
\boldsymbol A ^\textrm{H} \hat{\boldsymbol R}_0 ^{-\frac{1}{2}}
\end{equation}
and $\hat{ \boldsymbol V}_s$ contains the dominant right singular
vectors of the matrix $\hat{ \boldsymbol \Phi} ^ {-\frac{1}{2}}
\hat{ \boldsymbol \Sigma}$, while the associated singular values
are gathered in matrix $\hat{ \boldsymbol \Phi}_s$.

Using the IV-SSF method, the asymptotic error covariance matrix
has been shown to be \cite{viberg}
\begin{eqnarray}
\label{eq::bound_IV_SSF} \textrm{Cov}^\textrm{IV-SSF} =
\frac{1}{2n} \Big[ \Re \big(\boldsymbol D^\textrm{H} \boldsymbol C
^{-\frac{1}{2}} {\boldsymbol \Pi}_0^\bot \boldsymbol
C ^{-\frac{1}{2}} \boldsymbol D \big) \odot \qquad \quad \nonumber \\
\big( \boldsymbol {\mathcal J} ^{\, \textrm{H}}  \boldsymbol
{\mathcal A}_M ^\textrm{H} \, \boldsymbol \Phi^{-1}  \boldsymbol
{\mathcal A}_M \boldsymbol {\mathcal J} \big) ^ \textrm{T} \Big] ^
{-1}
\end{eqnarray}
in which $\boldsymbol {\mathcal A}_M = \boldsymbol I_M \otimes
\boldsymbol A$. The covariance in \eqref{eq::bound_IV_SSF} is
slightly different from the $\textrm{CRB}^\textrm{cor}$ in
\eqref{eq::CRB_cor_simp}. We aim to show that the IV-SSF method
with above error covariance does not attain the $\textrm{CRB}
^\textrm{cor}$ asymptotically and relying on this analysis, we
propose a variation in the method to boost its performance. To
this end, we use an asymptotic analysis presented in \cite{viberg}
with some modifications to be able to compare the $\textrm{CRB}^
\textrm{cor}$ and the covariance in \eqref{eq::bound_IV_SSF}.
Define an estimation problem in which we are interested to
estimate the signal $\boldsymbol s(t)$ using the data $\boldsymbol
z(t) = \boldsymbol s(t) + \boldsymbol w(t)$, in which $\boldsymbol
w(t)$ is a temporally white noise term with spatial covariance
matrix $(\boldsymbol A ^\textrm{H} \boldsymbol C ^{-1} \boldsymbol
A ) ^{-1}$. Note that $\boldsymbol w(t)$ is temporally white since
it is constructed from iid noise term $\boldsymbol \nu (t)$ by a
linear transform in the space domain. Assume that we are to
estimate $\boldsymbol s(t)$ from $M$ previous samples of
$\boldsymbol z(t)$
\begin{equation}
\label{eq::z} \boldsymbol z_M(t) = \big[ \, \boldsymbol
z^\textrm{T} (t-1) \quad \cdots \quad \boldsymbol z^\textrm{T}
(t-M) \, \big] ^\textrm{T}.
\end{equation}
with the best linear transform which minimizes the error
covariance matrix. The estimation error will be
\begin{equation}
\boldsymbol s(t) - \boldsymbol {{\mathcal H} z}_M(t) = \boldsymbol
e(t)
\end{equation}
where $\boldsymbol {\mathcal H}$ is the best linear transform in
the least squares sense. Making use of the orthogonality
principle, we can show that
\begin{equation}
\hat{ \boldsymbol {\mathcal H}} = \boldsymbol {\mathcal J}
^\textrm{H} \big( \boldsymbol {\mathcal P}_M + (\boldsymbol
{\mathcal A}_M ^\textrm{H} \boldsymbol {\mathcal C}_M ^{-1}
\boldsymbol {\mathcal A}_M ) ^{-1} \big)^{-1}
\end{equation}
where $\boldsymbol {\mathcal P}_M$ is the space-time covariance
matrix of the signals in $M$ snapshots and $\boldsymbol {\mathcal
C}_M = \boldsymbol I_M \otimes \boldsymbol C$. The resulting
minimized error covariance matrix will be
\begin{equation}
\boldsymbol \Sigma_e = \boldsymbol P_0 - \boldsymbol {\mathcal J}
^\textrm{H} \big( \boldsymbol {\mathcal P}_M + (\boldsymbol
{\mathcal A}_M ^\textrm{H} \boldsymbol {\mathcal C}_M ^{-1}
\boldsymbol {\mathcal A}_M ) ^{-1} \big) ^{-1} \boldsymbol
{\mathcal J}
\end{equation}
which along with a readily shown result similar to
\eqref{eq::pr_18} reduces the second part of the error covariance
matrix of IV-SSF method in \eqref{eq::bound_IV_SSF} to
\begin{eqnarray}
\label{eq::app_perf_IVSSF} \boldsymbol {\mathcal J} ^\textrm{H}
\boldsymbol {\mathcal A}_M ^\textrm{H} \, \boldsymbol \Phi^{-1}
\boldsymbol {\mathcal A}_M \, \boldsymbol {\mathcal J} =
\boldsymbol {\mathcal J} ^\textrm{H}
\big( \boldsymbol {\mathcal P}_M +  \qquad \qquad \nonumber \\
(\boldsymbol {\mathcal A}_M ^\textrm{H} \boldsymbol {\mathcal C}_M
^{-1} \boldsymbol {\mathcal A}_M ) ^{-1} \big) ^{-1} \boldsymbol
{\mathcal J} = \boldsymbol P_0 - \boldsymbol \Sigma_e.
\end{eqnarray}
The error covariance matrix in \eqref{eq::app_perf_IVSSF} has the
implication that the minimum attainable error covariance in the
IV-SSF method depends on the minimum error of prediction of the
process $\boldsymbol s(t)$ in the noise term $\boldsymbol w(t)$
using $M$ previous samples of the data $\boldsymbol z(t)$.
Lowering the prediction error will cause the error covariance of
IV-SSF in \eqref{eq::bound_IV_SSF} to reduce. Now we use an
asymptotic analysis on the $\textrm{CRB} ^\textrm{cor}$ to reform
it in a form similar to \eqref{eq::app_perf_IVSSF} which enables
us to understand why IV-SSF method does not reach the
$\textrm{CRB} ^\textrm{cor}$ and propose a modification in the
method to improve its performance. Assume that large number of
data is available $n \to +\infty$. The block trace in the
$\textrm{CRB} ^\textrm{cor}$ in \eqref{eq::CRB_cor_simp} can be
written as
\begin{equation}
\label{eq::BTr_expan} \textrm{BTr}_m [ \boldsymbol
{\mathcal{PA}}^\textrm{H} \boldsymbol {\, \mathcal{R}} ^ {-1}
\boldsymbol {\! \mathcal{A \, P}} ] = \sum_{i=1}^{n} \boldsymbol
{\mathcal P}^{i \cdot} \boldsymbol {\! \mathcal A} ^\textrm{H}
\boldsymbol {\, \mathcal{R}} ^ {-1} \boldsymbol {\! \mathcal{A \,
P}} ^{\cdot i}
\end{equation}
where $\boldsymbol {\mathcal P}^{i \cdot}$ and $\boldsymbol {
\mathcal P} ^{\cdot i}$ denote the $i$'th block row and column of
$\boldsymbol {\mathcal P}$, respectively. We assume a stationary
signal model in this section which results in a block Toeplitz
signal space-time covariance $\boldsymbol {\mathcal P}$. Further,
we assume a regular signal random process in which the
cross-covariance decreases with increasing time lag. For such
asymptotic conditions we can see that block rows $\boldsymbol
{\mathcal P}^{i \cdot}$ are shifted versions of each other,
ignoring the first and last ones. The same is true for block
columns $\boldsymbol { \mathcal P} ^{\cdot i}$ and also for the
matrix $\boldsymbol {\mathcal R}^{-1}$ and hence $\boldsymbol {
\mathcal A} ^\textrm{H} \boldsymbol {\, \mathcal{R}} ^ {-1}
\boldsymbol {\! \mathcal A}$, i.e. block rows (and columns) of
these matrices are shifted versions of each other. This special
matrix multiplication form ensures that the terms in summation in
\eqref{eq::BTr_expan} are approximately equal. Then,
\eqref{eq::BTr_expan} can be reduced to
\begin{equation}
\textrm{BTr}_m [ \boldsymbol {\mathcal{PA}}^\textrm{H} \boldsymbol
{\, \mathcal{R}} ^ {-1} \boldsymbol {\! \mathcal{A \, P}} ] \simeq
n \boldsymbol {\mathcal P'} ^{\textrm{T}} \boldsymbol {\! \mathcal
A} ^\textrm{H} \boldsymbol {\, \mathcal{R}} ^ {-1} \boldsymbol {\!
\mathcal{A \, P'}}
\end{equation}
where
\begin{equation}
\label{eq::P'} \boldsymbol {\mathcal P'} = [ \; \cdots \quad
\boldsymbol P_{-1} ^\textrm{T} \quad \boldsymbol P_0 ^\textrm{T}
\quad \boldsymbol P_{+1} ^\textrm{T} \quad \cdots \; ]
^\textrm{T}.
\end{equation}
Now, using a similar analysis as in \eqref{eq::z} to
\eqref{eq::app_perf_IVSSF} with a data vector defined as
\begin{eqnarray}
\label{eq::z_CRB} \boldsymbol z_{crb}(t) = \big[ \cdots
\boldsymbol z^\textrm{T} (t-1) \quad \boldsymbol z^\textrm{T} (t)
\quad \boldsymbol z^\textrm{T} (t+1) \cdots \big] ^\textrm{T}
\end{eqnarray}
results in an approximation similar to \eqref{eq::app_perf_IVSSF}
for the second part of the $\textrm{CRB} ^ \textrm{cor}$
\begin{equation}
\boldsymbol {\mathcal P'} \boldsymbol {\! \mathcal A} ^\textrm{H}
\boldsymbol {\, \mathcal{R}} ^ {-1} \boldsymbol {\! \mathcal{A \,
P'}} = \boldsymbol P_0 - \boldsymbol \Sigma_\epsilon
\end{equation}
in which, $\boldsymbol \Sigma_\epsilon$ is the error covariance of
the \emph{smoothing} of the random process $\boldsymbol z(t)$,
i.e. estimating the signal part $\boldsymbol s(t)$ from the data
described in \eqref{eq::z_CRB}. Now, after that we have
transformed the matrix forms of the $\textrm{CRB} ^ \textrm{cor}$
and the error of the IV-SSF method to the same formats, it is
possible to see that the IV-SSF method won't attain the
$\textrm{CRB} ^ \textrm{cor}$ since $\boldsymbol \Sigma_\epsilon <
\boldsymbol \Sigma_e$. Obviously, it is easier and results in
lower error to estimate the signal $\boldsymbol s(t)$ using
$\boldsymbol z_{crb}(t)$ in \eqref{eq::z_CRB} rather than using
$\boldsymbol z_M(t)$ in \eqref{eq::z} because of larger amount of
correlated data available. Using this analysis, we can understand
what is required to improve the performance of the IV-SSF method.
The key observation is that the structure of the estimation
problem and $\boldsymbol z_M(t)$ is similar to the structure of
the $\boldsymbol \phi(t)$, the instrumental variable chosen for
the IV-SSF. It shows that we could achieve the best performance if
we could have defined an instrumental variable containing all data
before and after the present signal $\boldsymbol s(t)$. Though, we
can't include $\boldsymbol x(t)$ in the instrumental variables
since it requires the knowledge of the noise covariance matrix.
Also, the limited degree of complexity we afford for our method
don't permit the inclusion of too many instrumental variables. We
are free to choose $M$ instrumental variables and we propose the
following
\begin{equation}
\boldsymbol \phi_{pro}(t) \triangleq \left[ \!\! \begin{array}{c}
\boldsymbol x(t+ \frac{M}{2}) \\ \vdots \\ \boldsymbol x(t+1) \\
\boldsymbol x(t-1) \\ \vdots \\ \boldsymbol x(t-\frac{M}{2})
\end{array} \!\! \right]
\end{equation}
In the subsequent section, we will present simulation results
which confirm the improvement in the performance of the IV-SSF
method using the above two-sided instrumental variables.

\section{Simulation Results}
\label{sec::simulation}

In this section, we illustrate the $\textrm{CRB} ^\textrm{cor}$
and the performance of the two methods investigated: the IV-SSF
method and our proposed method which we call two-sided IV-SSF. We
also present simulation results that confirm the inequality in
theorem \ref{thrm::bound}. Performance of the methods is
considered in the sense of error variance and bias. Although the
IV-SSF method shows to be asymptotically ($n \to +\infty$)
unbiased and this is shown theoretically in \cite{viberg}, in the
nonasymptotic region of limited available data, it shows strong
bias. We will see that the two-sided IV-SSF outperforms IV-SSF in
both lower bias and lower error variance.

\begin{figure}
\centering
\includegraphics[width = 0.45\textwidth]{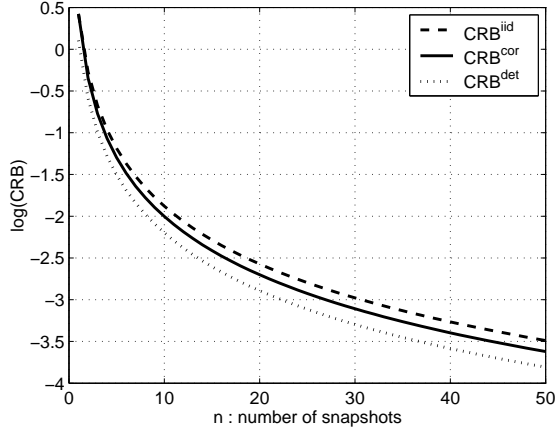}
\caption{An example of the CRB versus the number of snapshots in
logarithmic scale for two sources. The temporally correlated CRB
is upper bounded by the iid CRB and lower bounded by the
deterministic CRB.} \label{fig::CRB_n}
\end{figure}

\begin{figure}
\centering
\includegraphics[width = 0.45\textwidth]{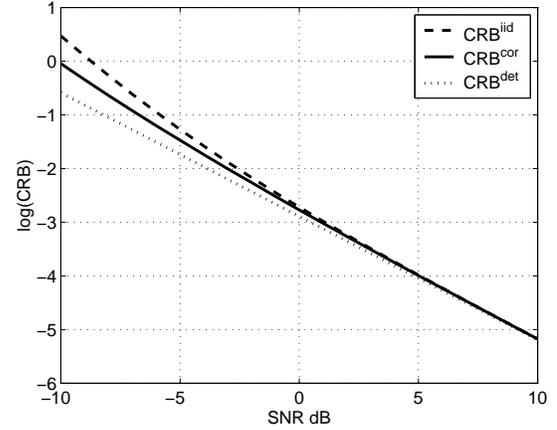}
\caption{An example of the CRB versus the SNR in logarithmic scale
for two sources. The temporally correlated CRB tends to the iid
CRB as SNR increases.} \label{fig::CRB_snr}
\end{figure}

\begin{figure}
\centering
\includegraphics[width = 0.45\textwidth]{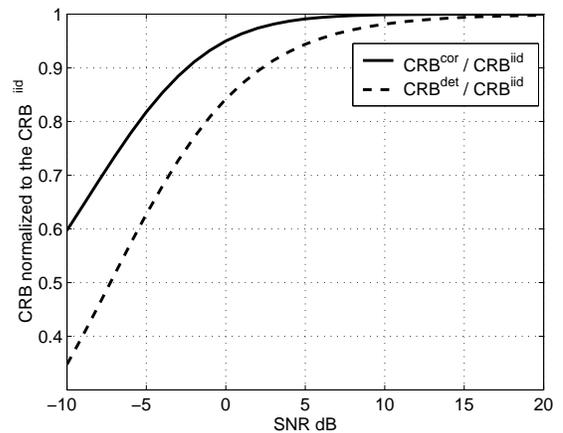}
\caption{The comparison of the temporally correlated CRB and
deterministic CRB normalized to the iid CRB. The comparison shows
the equivalence of the iid and temporally correlated CRBs when SNR
is high enough.} \label{fig::CRB_rel}
\end{figure}

We first present some graphs on the Cramer-Rao bounds. The main
result we provided on the CRBs is the inequality in
\eqref{eq::thrm_CRB}. We also showed that the temporally
correlated CRB is decreasing with the number of snapshots $n$.
These facts are presented via an example of two-source scenario in
Fig. \ref{fig::CRB_n}. Conditions of the simulation are as
follows: two sources at angles $[0 \;\; 0.2]$ radians with respect
to the array broadside are present. Array number of elements is 3
with $\lambda / \, 2$ spacing and SNR=10 dB. Uniform linear array
is assumed and the signal covariance matrix is defined as
\begin{equation}
\boldsymbol {\mathcal P} = \boldsymbol P_t \otimes \boldsymbol P_s
\end{equation}
where $[\boldsymbol P_t ] _{ij} = e^{-0.2 |i-j|}$, $[\boldsymbol
P_s] _{ij} = e^{-0.5 |i-j|}$, and $\boldsymbol C_{ij} = \sigma^2
e^{-|i-j|}$. Fig. \ref{fig::CRB_n} shows the decreasing CRB versus
increasing $n$, it also confirms the inequality in
\eqref{eq::thrm_CRB} for this special case. It is also noteworthy
that the difference between three types of CRBs increases with $n$
since with increased number of snapshots, there are more room for
the temporal correlation to improve the DOA estimation
performance. In Fig. \ref{fig::CRB_snr}, CRBs are depicted versus
SNR. It can be seen that as SNR increases, three types of CRB
merge together and decrease linearly. Though, the $\textrm{CRB}
^\textrm{cor}$ joins the $\textrm{CRB} ^\textrm{iid}$ faster than
$\textrm{CRB} ^\textrm{det}$. This is better shown in Fig.
\ref{fig::CRB_rel}, where the CRBs are normalized to the
$\textrm{CRB} ^\textrm{iid}$ to confirm the result of theorem
\ref{thrm::snr} which is stated in \eqref{eq::thrm_snr_high}.

\begin{figure}
\centering
\includegraphics[width = 0.45\textwidth]{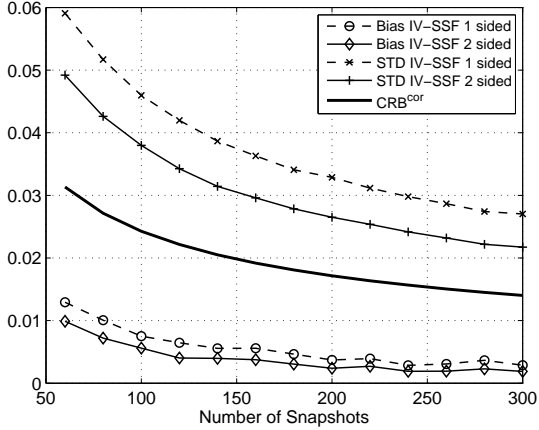}
\caption{Bias and standard deviation of one- and two-sided IV-SSF
methods versus $n$. It is clear that the two-sided IV-SSF
outperforms the one-sided one in both lower bias and lower
standard deviation. The improvement is rather constant with $n$.
Asymptotically in large $n$, bias is negligible in the overall
mean square error.} \label{fig::bias_std_crb_n}
\end{figure}

Now, we present simulation results which support the proposed
method versus the main IV-SSF. In the simulations, single source
located in $\omega \triangleq 2\pi \frac{d}{\lambda} \cos( \theta)
= 0.8$ is assumed, with a four-element half-wavelength array,
number of instrumental variables $M=2$, and number of trials
10000. In Fig. \ref{fig::bias_std_crb_n}, performance measures,
bias and standard deviation of the estimate of $\omega$ are
presented versus the number of snapshots $n$, while SNR=0 dB is
constant. Signal temporal correlation is simulated via filtering
an iid random sequence with an FIR filter with relative tap
weights
\begin{equation}
\label{eq::f(z)} f(z) = 1 + 0.5z^{-1} + 0.3z^{-2} + 0.2z^{-3} +
0.1z^{-4}
\end{equation}
which is then normalized to give a unit-energy filter. The
estimates are calculated using a two step grid search, first a
coarse search with grid size 0.01 and then a finer one with grid
size 0.001. Fig. \ref{fig::bias_std_crb_n} shows the bias and
standard deviation of both methods versus $n$. Although bias is
relatively small, it cannot be neglected in small numbers of
snapshots. Although the $\textrm{CRB}^{\textrm{cor}}$ in
\eqref{eq::CRB_cor_simp} has been calculated assuming zero bias,
it is still rewarding to compare the performance of the methods to
the square root of the $\textrm{CRB}^{\textrm{cor}}$, which is
depicted in Fig. \ref{fig::bias_std_crb_n}. Here, we used $f(z)$
in \eqref{eq::f(z)} to compute the signal space-time covariance
matrix $\boldsymbol {\mathcal {P}}$. The improvement made by the
two-sided IV-SSF is rather constant with $n$ specially in the
standard deviation. It is clear that the two-sided IV-SSF had made
roughly 20\% improvement in the standard deviation and bias
without increasing the computational load of the algorithm.

\begin{figure}
\centering
\includegraphics[width = 0.45\textwidth]{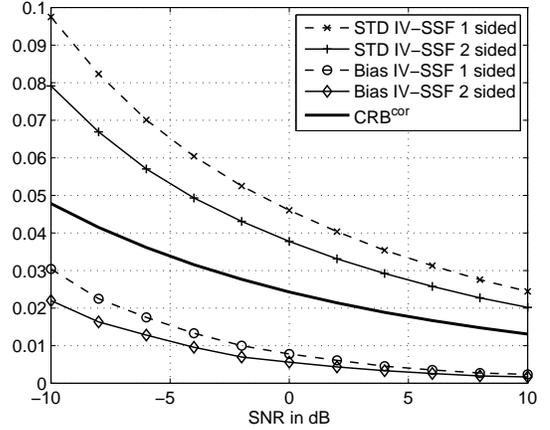}
\caption{Bias and standard deviation of one- and two-sided IV-SSF
methods versus the SNR. The improvement made by two-sided IV-SSF
is larger in lower SNRs. In high SNRs, bias is negligible in the
overall error.} \label{fig::bias_std_crb_snr}
\end{figure}

\begin{figure}
\centering
\includegraphics[width = 0.45\textwidth]{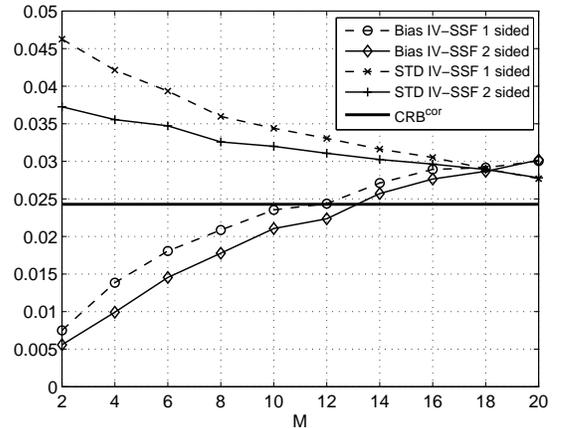}
\caption{Bias and standard deviation of one- and two-sided IV-SSF
methods versus the number of instrumental variables $M$. The
improvement made by two sided IV-SSF in standard deviation is
larger in low M. Also, bias grows with M and finally gets larger
than standard deviation for large M.} \label{fig::bias_std_crb_M}
\end{figure}

In Fig. \ref{fig::bias_std_crb_snr}, the same comparison is made
with constant $n = 100$ and varying SNR. Obviously, the
improvement made by two-sided IV-SSF is greater in lower SNR. In
high SNR conditions, bias is rather negligible in both methods,
while the improvement percentage is nearly constant with SNR.
Finally in Fig. \ref{fig::bias_std_crb_M}, the comparison is made
in constant $n=100$ and SNR=0 dB, while $M$ is changed to see the
effect of the method complexity indicator or the number of
instrumental variables on the bias and standard deviation of both
methods. The interesting point is that in low $M$, there is
considerable improvement from one-sided to two-sided IV-SSF. But
this improvement decreases as $M$ increases since for large $M$,
there is enough previous samples to estimate the present signal of
interest with sufficient accuracy comparable to the accuracy that
can be achieved using the two-sided estimation. Another important
point of the Fig. \ref{fig::bias_std_crb_M} is the increasing
relative part of the bias in the overall mean square error of the
estimation with increasing $M$. This phenomena can lead us to
limit the number of instrumental variables so as to avoid large
unpredictable bias. This selection can be more rational when we
consider the increased cost of implementation when $M$ gets
larger. It is also evident in Fig. \ref{fig::bias_std_crb_M} that
the $\textrm{CRB}^{\textrm{cor}}$ is independent of $M$ and
gradually, with increasing $M$, the portion of error caused by
standard deviation decreases while the error caused by bias
increases.

\section{Conclusion}
\label{sec::Conclusion}

In this paper, we presented new theoretical results on the
performance of the DOA estimation when signals of interest are
possibly temporally correlated. In particular, it was shown that
the Cramer-Rao bound in the temporally correlated signal model is
upper bounded by the same bound under the iid signal model. This
result implies that temporal correlation of the signals is an
additional relevant information for DOA estimation. We showed that
the improvement caused by signal temporal correlation is large
when SNR is low and little when SNR is high. This is a good news
since the high SNR condition is not critical in our systems where
we can use many suboptimal methods; while the low SNR performance
improvement is valuable to the system. The second part of the
paper was devoted to a practical method of DOA estimation in the
new signal model. The IV-SSF method was analyzed and compared with
the $\textrm{CRB} ^\textrm{cor}$ to show that it is not an
efficient method. Then, using a special form of the $\textrm{CRB}
^\textrm{cor}$, we proposed a version of the IV-SSF method
(two-sided IV-SSF), that outperform the former method in both
lower bias and lower error variance.

\end{document}